# Time resolved photoluminescence method: determination of the recombination mode

Pawel Strak[1], Kamil Koronski[2], Konrad Sakowski[1], Kamil Sobczak[3], Jolanta Borysiuk[4], Krzysztof P. Korona[4], Andrzej Suchocki[2], Eva Monroy[5], Stanislaw Krukowski[1], and Agata Kaminska[2,6, *]

[1]*Institute of High Pressure Physics, Polish Academy of Sciences, Sokolowska 29/37, 01-142 Warsaw, Poland*

[2]*Institute of Physics, Polish Academy of Sciences, Aleja Lotnikow 32/46, PL-02668 Warsaw, Poland*

[3]*Faculty of Chemistry, Biological and Chemical Research Centre, University of Warsaw, Zwirki i Wigury 101, 02-089 Warsaw, Poland*

[4]*Faculty of Physics, University of Warsaw, Pasteura 5, 02-093 Warsaw, Poland*

[5]*Université Grenoble-Alpes, CEA, INAC-Pheliqs, 17 av. des Martyrs, 38000 Grenoble, France*

[6] *Cardinal Stefan Wyszynski University, Faculty of Mathematics and Natural Sciences. School of Exact Sciences, Dewajtis 5, 01-815 Warsaw, Poland*

*Corresponding author: +48 22 116 3519; fax +48 22 847 5223

*E-mail address:* kaminska@ifpan.edu.pl (A. Kaminska).

## Abstract

A method of data analysis is proposed for the determination of the carrier recombination processes in optically excited matter measured by time-resolved photoluminescence (PL), differentiating monomolecular, bi-molecular, tri-molecular, and higher order molecular processes. The procedure is applicable to the time evolution of any optically excited system. The method is based on the introduction of instantaneous PL recombination rate $r_{PL}$ plotted as a function of the PL intensity or the time. The method provides deep insight into the time evolution of the recombination of the optically excited systems. As an illustration, the method is successfully applied to III-nitride polar and non-polar multi-quantum wells (MQWs). At low temperatures (5K), the mono- and bi-molecular processes determine the carrier relaxation, and the tri-molecular Auger recombination contribution is negligible. At room temperature, the data indicate an important contribution of Auger processes. It is also shown that asymptotic (low excitation), linear recombination rate has different character: a considerable temperature dependent contribution to overall recombination rate exists for polar MQWs while for nonpolar MQWs the rate does not depend on the temperature.

# 1. Introduction

Photoluminescence and its time-resolved version (TRPL) are keystone characterization techniques to explore properties of systems as diverse as single atoms, molecules, solids, and a variety of quantum-confined structures, such as quantum wells, rods and dots. PL measurements probe the electronic states of the system [1,2]. As the amount of potential information is enormous, the PL measurement techniques steadily progress, extending their scope and application range through different modifications [3]. Among the adopted features, TRPL is one of the most important and productive [3,4]. In TRPL measurements, optical excitation induces nonequilibrium occupation of a portion of quantum states, which then relax towards equilibrium. For extended systems, the nonequilibrium density of electrons and holes (e-h) is created by an initial laser pulse, and the excess of carriers relaxes by recombination. Part of the recombination processes entails emission of photons (i.e. radiative recombination) which are measured by optical detectors. The deexcitation processes without emission of photons contribute to non-radiative recombination. All the types of recombination may involve one, two, or more carriers; they are called mono-, bi-,and tri-molecular processes. The time evolution of the excess carrier density may be studied by measurements of time evolution of the PL intensity. Solutions exist for each separate molecular type process, thus single processes can be distinguished easily [4]. In reality, several types of recombination can be entangled simultaneously and their deconvolution is difficult. Thus the direct and unambiguous PL time evolution analysis method does not exist in practice.

In practical analysis of the experimental results, it is assumed that the time decay of the optical emission of an excited semiconductor medium can be divided into monomolecular, bi-molecular, tri-molecular and possibly higher-order molecular processes [1-5]. It has to be also noted that division of the recombination channels into mono-, bi-, tri-, and higher molecular recombination modes does not entail precise identification of the nature of the recombination process. Thus, the monomolecular decay may contain contributions from several recombination modes, including non-radiative recombination, related to presence of defects, known as Shockley-Read-Hall (SRH) recombination [6], and also Wannier-Mott exciton recombination [7,8]. The contribution of single and collective excitons is dominant for low temperatures [9], i.e. at the conditions of majority of PL experiments. However, for high electron densities, excitons can be screened or even decomposed, so that they do not play a dominant role [10,11]. Additionally, for recombination in multiquantum wells (MQWs), the escape from the wells may contribute to monomolecular decay [12]. Several physical mechanisms may lead to recombination outside QWs, including lack of electron capture in the wells, electron escape, asymmetry of electron and hole transport in semiconductors, etc [13]. In many cases, additional experiments such as temperature or pressure dependent PL, in combination with *ab initio* calculations [14], are necessary to disentangle the different contributions to mono-molecular processes.

The radiative recombination, especially in semiconducting materials, may be related to band-to-band recombination, direct or indirect, which may be associated either with mono-molecular or bi-molecular recombination, depending on the condition of the experiment [15]. The observed PL time decay may be additionally affected by the energy transfer by exchange mechanisms, involving defects [16,17]. For example, the peculiar behavior of the PL intensity from $In_xGa_{1-x}N$ ($0 < x < 0.25$) layers and MQWs was explained by Chichibu's localization model, according to which, the carriers are localized in high In content regions, and isolated from dislocations, which constitute the non-radiative recombination channels [18,19]. As it was confirmed experimentally [20,21], the localization works at low excitation, hence it is expected that at higher excitation, during the time evolution, the transition from non-localized to localized recombination mode may alter the results. In addition, the built-in and piezoelectric-induced electric fields in the wells [21-24] may affect the recombination by at least two different mechanisms: field-induced dissociation of the excitons and energy shift due to the screening of the electric field by photogenerated carriers [25].

The most prominent example of a tri-molecular or higher molecular process is Auger recombination. The process involves non-radiative transfer of the excess energy created due to e-h recombination to a third charge carrier (electron or hole), which is ejected from its original state [26,27]. Depending on the final state of the ejected carrier, several types of Auger processes were identified, e.g. internal, external. Auger recombination was recently identified as the most likely phenomenon responsible for the efficiency droop in III-nitride light emitting diodes (LEDs) [28,29], i.e. the relative decrease of light emission under high injection current [30,31]. However, the Auger coefficients are hard to determine, since measurements are burdened by the influence of the e-h ratio, electric field, or hot carriers escape [32]. As a result, the Auger coefficients may vary by one or even two orders of magnitude [29]. Thus, there is an urgent demand of precise, independent assessment of the contribution of Auger processes to e-h recombination, and the determination of the Auger coefficients.

The purpose of this work is to present a new method of analysis of the temporal evolution of the PL intensity, based on a logarithmic derivative method. The method allows the identification of the mono-, bi- and tri- molecular, and higher order PL decay processes. Examples of application are shown, demonstrating the power of the method. The perspectives of application in basic research and for the characterization of structures and devices are described.

## 2. Theoretical background

Time-resolved PL experiments proceed via excitation of the semiconductor sample by an intense laser pulse. The energy of the laser radiation is such that a photon can generate a single electron-hole (e-h) pair in the conduction and the valence band. The time dependent excess densities of electrons and holes are denoted as $n(t)$ and $p(t)$, respectively. At the end of excitation pulse, i.e. the time $t_o$, these densities are:

$$n(t_o) = n_{tot}(t_o) - n_{eq} \qquad (1a)$$

$$p(t_o) = p_{tot}(t_o) - p_{eq} \qquad (1b)$$

where the sufixes 'tot' and 'eq' denote total and equilibrium density of the carriers, respectively. In the case of n-type structure, like in the example considered below, $n_{eq} >> p_{eq}$. We assume that each photon generates an e-h pair, so that $n(t_o) = p(t_o)$. After the pulse, during the PL emission process ($t > t_o$), the e-h pairs are annihilated due to recombination. Therefore, for negligible carrier escape, we can write $n(t) = p(t)$ for $t > t_o$.

The main *radiative recombination*, which generates the PL signal, proceeds by band-to-band transitions involving an e-h pair, i.e. it is a bi-molecular process. Thus, the radiative recombination rate reflects the bi-molecular character of the process:

$$R_b(t) = B n_{tot}(t) p_{tot}(t) \qquad (2)$$

where $B$ ($B > 0$) is a proportionality constant.

On the other hand, the recombination rate may also include radiative excitonic transitions and/or non-radiative *SRH recombination* related to the presence of the defects, when photogenerated carriers are captured at the defect site and then they gradually relax towards the opposite band (ground state) by emission of phonons. These are monomolecular processes described by recombination rate proportional to the excess carrier density, i.e.

$$R_m(t) = An_{tot}(t) \tag{3a}$$

or

$$R_m(t) = Ap_{tot}(t) \tag{3b}$$

where $A$ $(A > 0)$ is a proportionality constant associated to monomolecular process. Non-radiative SRH recombination is highly detrimental for the performance of optoelectronic devices, as it converts the absorbed energy into heat.

The third important recombination path is *Auger recombination*, a tri-molecular process involving a photogenerated e-h pair and a third carrier (electron or hole) carrying out the excess energy. Therefore, the Auger recombination rate is given by

$$R_t(t) = C_1 n_{tot}^2(t) p_{tot}(t) + C_2 p_{tot}^2(t) n_{tot}(t) \tag{4}$$

Where $C_1, (C_1 > 0)$ and $C_2, (C_2 > 0)$ are proportionality constants determined by the dynamics of Auger processes, corresponding to non-radiative energy transfer to electron and hole, respectively.

These three recombination paths contribute to the decrease of the excess carrier density, $n(t)$ and $p(t)$. Under high excitation, the excess densities are much higher than the equilibrium densities of both types of carriers. Therefore, for some initial period ($t > t_o$),

$$n_{tot}(t) \cong n(t) \tag{5a}$$

$$p_{tot}(t) \cong p(t) = n(t) \tag{5b}$$

In this case, the fundamental rate equation describing the optical relaxation, known as ABC equation [33-36], is

$$-\frac{dn}{dt} = R_m + R_b + R_t = An + Bn^2 + Cn^3 \qquad (6)$$

The measured PL intensity $J_{PL}^b(t)$ is proportional to the radiative recombination rate $R_b(t)$ and the geometric factor of the photodetector $\Gamma$:

$$J_{PL}^b(t) = \Gamma R_b(t) = \Gamma B n^2(t) \qquad (7)$$

As the relevant information is contained in the time evolution, the PL intensity can be normalized to unity at the initial time $t_o$, i.e. $J_{PL}^b(t_o) = B n_o^2 \Gamma$ where the initial carrier density $n(t_o) = n_o$. Accordingly, Eq. (7) can be written as:

$$J_{PL}^b(t) = \left(\frac{n(t)}{n_o}\right)^2 \qquad (8)$$

By division of Eq. (6) by initial carrier density $n_o$:

$$-\frac{d}{dt}\left(\frac{n}{n_o}\right) = A\frac{n}{n_o} + Bn_o\left(\frac{n}{n_o}\right)^2 + Cn_o^2\left(\frac{n}{n_o}\right)^3 \qquad (9)$$

Multiplying both sides of the equation by $2\,n(t)/n_o$ leads to the following expression for the time derivative of the PL intensity:

$$-2\frac{n}{n_o}\frac{d}{dt}\left(\frac{n}{n_o}\right) = -\frac{d}{dt}\left(\frac{n}{n_o}\right)^2 = -\frac{dJ_{PL}^b}{dt} = 2\frac{n}{n_o}\left[A\frac{n}{n_o} + Bn_o\left(\frac{n}{n_o}\right)^2 + Cn_o^2\left(\frac{n}{n_o}\right)^3\right] \qquad (10)$$

Finally, the division by PL intensity $J_{PL}^b$ gives the final result:

$$-\frac{1}{J_{PL}^b}\frac{dJ_{PL}^b}{dt} = -\frac{d}{dt}\left[ln\left(J_{PL}^b\right)\right] = 2A + 2Bn_o\sqrt{J_{PL}^b} + 2Cn_o^2 J_{PL}^b \qquad (11)$$

which allows a determination of the various contributions to carrier relaxation based on the analysis of the polynomial-form dependence of the logarithmic derivative, as it will be shown later.

An additional insight into the nature of optical relaxation may be gained by generalization of the simplest single exponential decay of the PL intensity, given as [4]:

$$J_{PL}^{exp}(t) = exp\left(-\frac{t}{\tau_{PL}}\right) \qquad , \qquad (12)$$

where $J_{PL}^{exp}(t)$ denotes normalized PL intensity. Then the standard PL recombination rate may be defined as the inverse of relaxation time:

$$r_{PL} = \frac{1}{\tau_{PL}} = -\frac{d}{dt}\left[ln\left(J_{PL}^{exp}(t)\right)\right] \tag{13}$$

This formulation may be generalized for the above discussed PL intensity time evolution $J_{PL}^{b}$ as follows:

$$r_{PL}(t) = \frac{1}{\tau_{PL}} = -\frac{d}{dt}\left[ln(J_{PL}^{b})\right] = 2A + 2Bn_o\sqrt{J_{PL}^{b}} + 2Cn_o^2 J_{PL}^{b} \tag{14}$$

where instantaneous PL recombination rate $r_{PL}$ depends on the time explicitly.

Note that Eqs. (11) and (14) are valid for high excitation only. Therefore, a deviation from the equation is expected for $t >> t_o$, i.e. when the excess carrier density decreases to a value comparable to equilibrium density of the dominant carrier type. The same situation arises when photogenerated carrier density is much lower than the density of majority carriers at equilibrium. In both cases the total density of majority carriers is approximately equal to equilibrium value, i.e. it is constant, independent of the time.

Under low excitation conditions, the total carrier density can be approximated by:

$$n_{tot}(t) \approx n_{eq} \qquad ; \qquad p_{tot}(t) \approx p(t) \tag{15a}$$

when the majority carriers are electrons, and

$$p_{tot}(t) \approx p_{eq} \qquad ; \qquad n_{tot}(t) \approx n(t) \tag{15b}$$

when the majority carriers are holes. Following Eq. (2), the radiative recombination rate can be approximated by

$$R_{rad}(t) = Bn_{tot}(t)p_{tot}(t) \approx Bp(t)n_{eq} \tag{16a}$$

or

$$R_{rad}(t) = Bn_{tot}(t)p_{tot}(t) \approx Bn(t)p_{eq} \tag{16b}$$

when the majority carriers are electrons or holes, respectively.

This means that the radiative recombination becomes monomolecular, governed by the excess density of the minority carriers. Note that for long enough time ($t \gg t_o$) the recombination process will always become monomolecular.

On the other hand, the SRH recombination contributes to the monomolecular recombination rate, keeping proportional to the carrier density, i.e. $R_m(t) = A p_{tot}(t)$ or $R_m(t) = A n_{tot}(t)$, for electrons or holes as majority carriers, respectively.

Finally, the tri-molecular process, Auger recombination, is described by:

$$R_t(t) = C_1 n_{tot}^2(t) p_{tot}(t) + C_2 p_{tot}^2(t) n_{tot}(t) \approx C_1 n_{eq}^2 p(t) + C_2 n_{eq} p^2(t) \qquad (17a)$$

when the majority carriers are electrons, and

$$R_t(t) = C_1 n_{tot}^2(t) p_{tot}(t) + C_2 p_{tot}^2(t) n_{tot}(t) \approx C_1 n^2(t) p_{eq} + C_2 n(t) p_{eq}^2 \qquad (17b)$$

when the majority carriers are holes. Hence, the Auger recombination adds to monomolecular rate, and to the bi-molecular recombination. Thus, for low excitation, Eq. (6) becomes, for electron as majority carriers:

$$-\frac{dp}{dt} = R_m + R_b + R_t = \left(A + B n_{eq} + C_1 n_{eq}^2\right) p + C_2 n_{eq} p^2 \qquad (18a)$$

and for holes:

$$-\frac{dn}{dt} = R_m + R_b + R_t = \left(A + B p_{eq} + C_2 p_{eq}^2\right) n + C_1 p_{eq} n^2 \qquad (18b)$$

As previously, the PL intensity $J_{PL}^m(t)$ is proportional to the radiative recombination rate $R_b(t)$ and the geometric factor $\Gamma$ (for holes as minority carriers):

$$J_{PL}^m(t) = \Gamma R_b(t) = \left[B n_{eq}\right] p(t) \qquad (19)$$

The geometric factor $\Gamma$ can be removed by normalizing PL intensity by its value at the initial time $t_0$:

$$J_{PL}^m(t) = \frac{p(t)}{p_0} \qquad (20)$$

Both sides of Eq. (18a) are divided by $p_o$ and by the PL intensity $J_{PL}^m(t)$ to arrive to the final expression:

$$-\frac{d}{dt}\left[ln\left(J_{PL}^m(t)\right)\right] = A_{eff} + B_{eff}J_{PL}^m(t) \tag{21a}$$

where

$$A_{eff} = A + Bn_{eq} + C_1 n_{eq}^2 \tag{21b}$$

and

$$B_{eff} = C_2 n_{eq} p_o. \tag{21c}$$

For electrons as minority carriers the expressions are obtained by replacement of the electrons and holes, respectively. Following Eq. (14), instantaneous PL relaxation rate $r_{PL}$ may be introduced as:

$$r_{PL}(t) = \frac{1}{\tau_{PL}} = -\frac{d}{dt}\left[ln(J_{PL}^m)\right] = A_{eff} + B_{eff}J_{PL}^m \tag{22}$$

which depends on the time explicitly.

Note the similarity of the derived expressions for high excitation and low excitation in Eqs (11) and (22), respectively. Therefore, it is possible to extract the coefficients associated with the different recombination processes from a common parabolic dependence:

$$F(y) = \alpha + \beta y + \gamma y^2 \tag{23}$$

where these factors depend on the type of the process:

High excitation: $y = \sqrt{J_{PL}^b} \quad \alpha = 2A \quad \beta = 2Bn_o \quad \gamma = 2Cn_o^2 \tag{24a}$

Low excitation: $y = J_{PL}^m \quad \alpha = A_{eff} \quad \beta = B_{eff} \quad \gamma = 0 \tag{24b}$

As shown in Fig. 1(a), the high excitation regime (labeled as a Process I) evolves towards low excitation (labeled as process II), characterized by linear dependence of PL intensity, i.e. leads to the situation when the relation from Eq. (7) is not fulfilled. Using the relation $J_{PL}^m = \left(\sqrt{J_{PL}^b}\right)^2$, the following dependence is obtained:

$$-\frac{d[ln(J_{PL}^b)]}{dt} = A + Bn_0\left(\sqrt{J_{PL}^b}\right)^2 \tag{25}$$

i.e. in the part of diagram corresponding to Process II, plotted in Fig. 1(a), the dependence on $\sqrt{J_{PL}^b}$ should be purely quadratic. Note that the transition between Process I and Process II type (vertical dotted line in Fig 1(a)) occurs when the photoexcited carrier density is approximately equal to the equilibrium density of majority carriers.

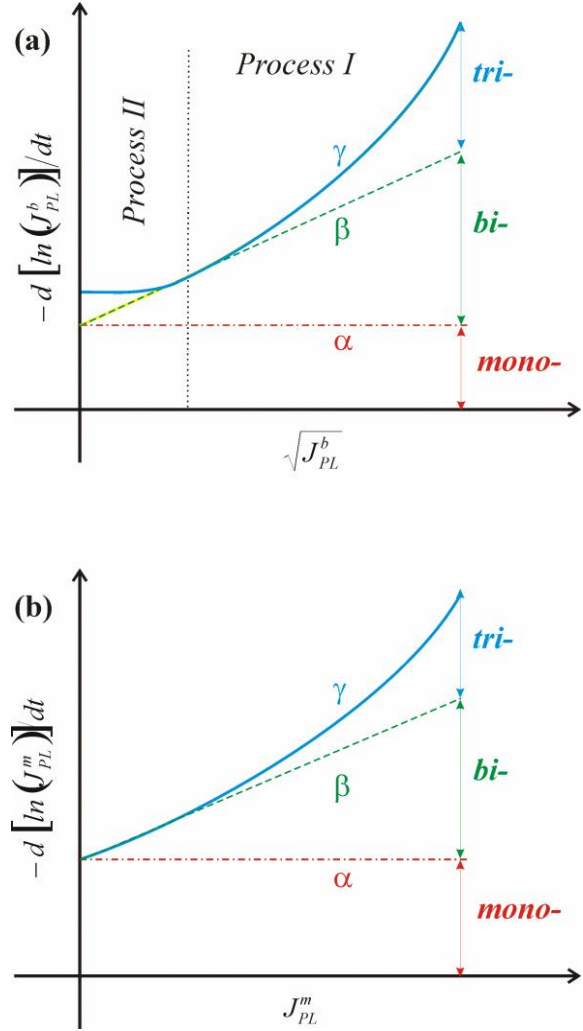

**Fig. 1.** Instantaneous PL relaxation rate $r_{PL}$, expressed as the logarithmic time derivative of PL intensity for two different excitation levels: (a) for high excitation, as a function of square root of PL intensity, and (b) for low excitation, as a function of PL intensity. In (a), the high excitation regime is separated from low excitation regime by vertical dotted line into process I (Eq. (11)) and process II (Eq. (21)), respectively. The red dash-dotted, green dashed and blue solid lines correspond to the evolution according to $\alpha$, $\alpha + \beta$ and $\alpha + \beta + \gamma$ parameters. The values of the parameters are: $\alpha = 1$, $\beta = 2$ and $\gamma = 3$. The relative weight of the various molecular processes at the initial level is marked by arrows for $J = 1$.

The implementation of the above presented procedure for determination of the various recombination type contributions and determination of the values of ABC parameters requires satisfying several stringent conditions. First, the averaged PL intensity should drop to zero for long enough times. Finally, it has to be stressed that the ABC model is a mere approximation of the more complex phenomena of optical relaxation, so that a deviation is likely to be detected. The possible limitation of the application of ABC model were discussed recently, showing that the presence of polarization induced field may affect Shockley – Read –Hall and Auger recombination rate [37,38]. This adds to the well-known effect of polarization induced fields in polar MQWs which reduces overlap of electron and hole wavefunctions and consequently the radiative recombination rates [39,40]. As it was shown, the radiative recombination rates may be reduced by three orders of magnitude by mere increase of the well width from 1 nm to 6 nm [41]. Thus, the ABC model needs to take into account influence of the polarization induced electric field as it may drastically affect the measured recombination rates.

It is also worth to note that application of logarithmic derivative was postulated earlier, but the analysis was based on the temporal dependence of carrier density n(t) [42]. The analysis was applied to time dependence of the signal, thus it could not provide data as precise as these derived by application of the above presented procedure.

## 3. Application to GaN/AlN multi-quantum-wells

### 3.1 Sample preparation

As an example, we present here the analysis of the TRPL from a polar GaN/AlN multi-quantum-well (MQW) structure grown on an AlN-on-sapphire substrate by plasma-assisted molecular beam epitaxy (PA-MBE) [23]. The MQW stack consists of 40 periods of 1.0-nm-thick GaN wells and 4-nm-thick AlN barriers. The GaN layers were doped with Si to a concentration of $1.3\times10^{19}$ cm$^{-3}$. A scheme of the sample and the scanning transmission electron microscopy (TEM) view of the MQW is presented in Fig. 2. The picture proves that the thicknesses of barrier/wells are almost uniform with low density of extended defects.

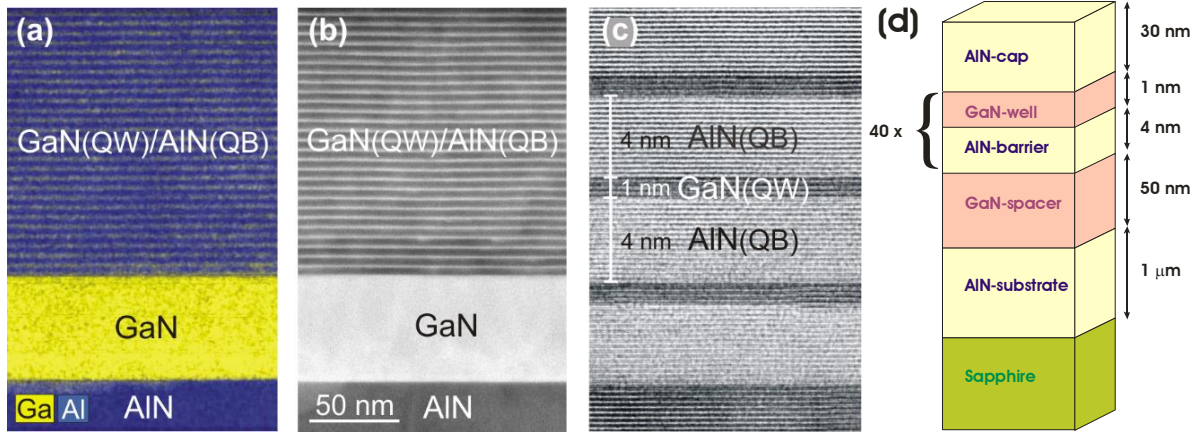

**Fig. 2.** Cross-sectional images of the GaN/AlN MQW structure with 1-nm-thick GaN wells and 4-nm-thick AlN barriers: (a) energy-dispersive X-ray spectroscopy (EDX) map of chemical composition of the structure, (b) high-angle annular dark-field - scanning transmission electron microscopy (HAADF-STEM) image, (c) high-resolution TEM (HR TEM) image, (d) scheme of the sample.

The choice of such narrow GaN wells aims at neglecting the effect of the polarization-related internal electric field on the evolution of the radiative recombination. In polar GaN/AlN heterostructures, the quantum-confined Stark effect (QCSE) leads to a red shift of the PL. Just after the excitation pulse, the electric field in polar QWs is partially screened by photoexcited carriers, what causes an initial increase of the emission energy (blue shift of the PL). Then, the recombination reduces screening, and the electric field is rebuilt resulting in a red shift as a function of time. Naturally, QCSE is less pronounced for narrow structures.

The use of non-polar crystallographic orientations allows for GaN/Al(Ga)N systems to operate without polarization-induced electric field [43], which simplifies the interpretation of TRPL data. As an example, we have characterized a non-polar m-plane MQW stack consisting of 50 periods of 3.0-nm-thick GaN wells and 22.5-nm-thick $Al_{0.26}Ga_{0.74}N$ barriers. The GaN layers were intentionally doped with Si to a concentration of 2 x $10^{19}$ cm$^{-3}$. The samples were grown by PAMBE at a substrate temperature $T_S$ = 720°C and with a nitrogen-limited growth rate of 0.4 ML/s ($\approx$ 360 nm/h). Growth was performed under slightly Ga-rich conditions [44,45]. The substrate was free-standing semi-insulating GaN sliced along the m-plane non-polar surface from (0001)-oriented GaN boules synthesized by hydride vapor phase epitaxy (resistivity >$10^6$ $\Omega$cm, dislocation density < $5\times10^6$ cm$^{-2}$).

## 3.2. TRPL measurements

Photoluminescence was excited with the third harmonic of a tunable Ti:sapphire pulsed laser Chameleon Ultra, Coherent Ltd. ($\lambda$ = 300 nm, $f$ = 80 MHz, pulse width 130 fs). The laser average power was up to 1 mW, impulse energy 12.5 pJ. The light beam was analyzed by spectrometer and directed on a CCD detector and a streak camera. The streak camera enabled time-resolved measurements, showing the time evolution of PL intensity and energy. In order to improve the signal-to-noise ratio, the PL decay data were taken as the average of more than $10^9$ decays.

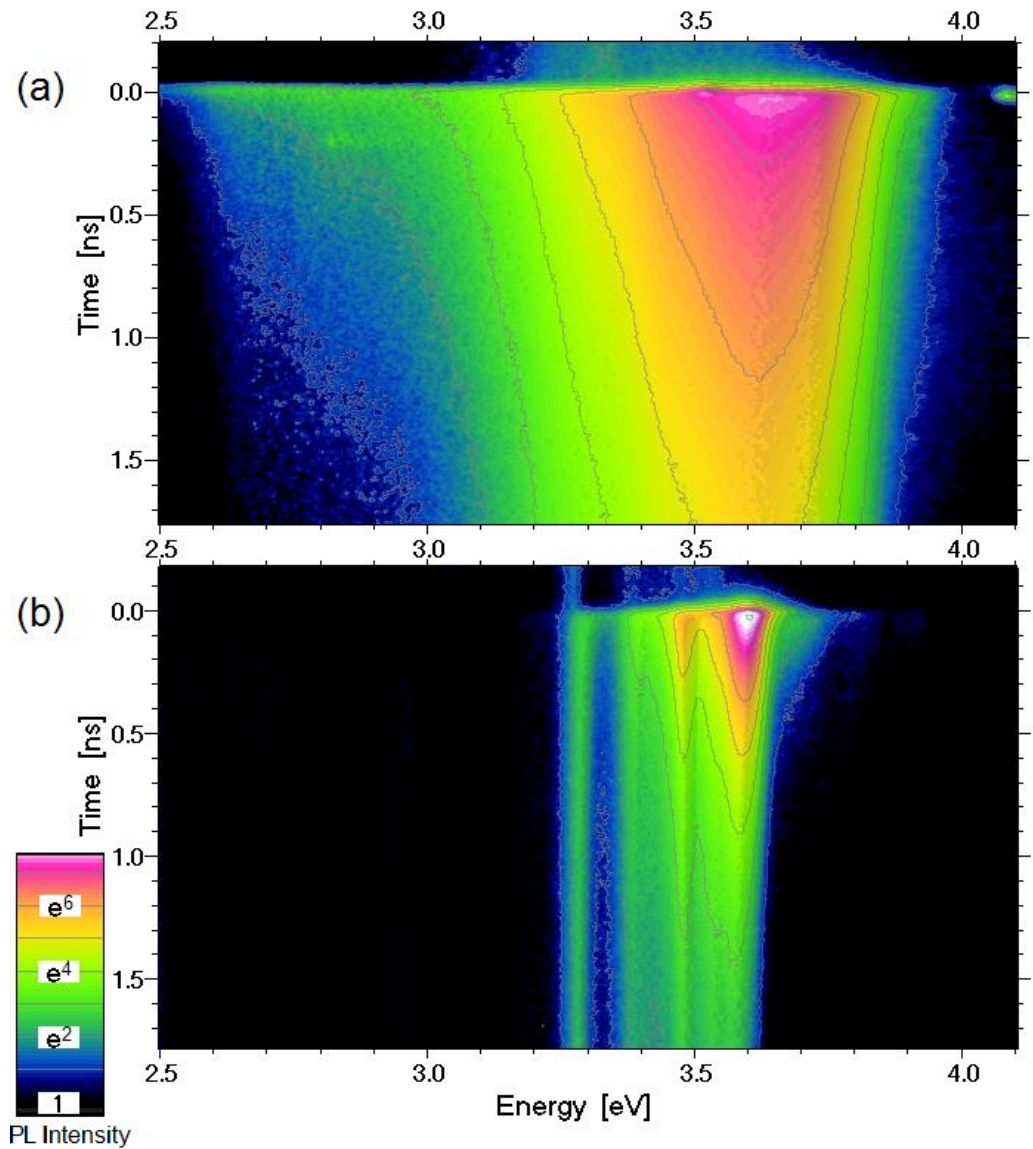

**Fig. 3.** Time-resolved PL spectra of (a) polar 1-nm-thick GaN/AlN MQWs and (b) non-polar 3-nm-thick GaN/Al$_{0.26}$Ga$_{0.74}$N MQWs. Contour lines are equally spaced in logarithmic scale (e – times spaced), so distance between two contour lines along time axis gives the decay time of the signal.

As shown in Fig. 3(a), the PL spectra from 1-nm-thick polar GaN/AlN QWs show no discernible energy shift in time. The energy is constant because the well is so narrow that the screening does not affect the quantum states during the recombination process. Similar time independent emission energy measured for the 3-nm-thick non-polar GaN/Al$_{0.26}$Ga$_{0.74}$N QWs is presented in Fig. 3(b).

### 3.3 Noise reduction

The logarithmic plot of PL intensity of polar nitride MQW structure versus time is shown in Fig. 4 (a) and (b) for room and liquid-helium temperatures, respectively. In addition to raw data, the data obtained by subtraction of the background value at infinity is shown. For long decay times, the PL decay is exponential.

The noise level was reduced by the following method: the smoothing interval was selected as a given odd number of points. Then according to least square method the linear approximation to the data in the interval was obtained. The central value was adopted as new value, replacing the old one. The procedure was repeated consecutively for all points along the line. At both ends, the data remain unchanged. As the noise to signal ratio was different for different times, the smoothing interval was different at both ends, equal to $N_{min}$ at the beginning, and $N_{max}$ at the end of PL decay. For the points between the smoothing interval length was scaled linearly. The procedure was repeated several times to obtain sufficient smoothing. For instance, for low temperature data (5K) obtained with an average laser power of 500 and 200 μW, the smoothing integral begins at t = 0 for 3 points and terminates at 121 and 81 points, respectively. The procedure was repeated 3 and 5 times respectively which was sufficient to remove the noise so that the derivative could be obtained [46]. It has to be stressed out that the number of points in smoothing interval and the number of iterations has to be drastically changed depending on the noise/signal ratio. As is clearly visible in Fig. 4, the proposed method of noise reduction does not distort results, and it is reasonable to assume that the function under consideration (represented by denoised data) is continuously differentiable. So, as a next step, from the smoothed time dependence, the logarithmic derivative of the PL intensity $-\frac{d[ln(J)]}{dt}$ as a function of the intensity $J$ was obtained. It has to be noted that the initial time dependence does not obey ABC dynamics, evidently due to electron and hole transportation/equilibration processes. Therefore the plotted diagram corresponds to the terminating phase of the evolution where the system remains in local equilibrium.

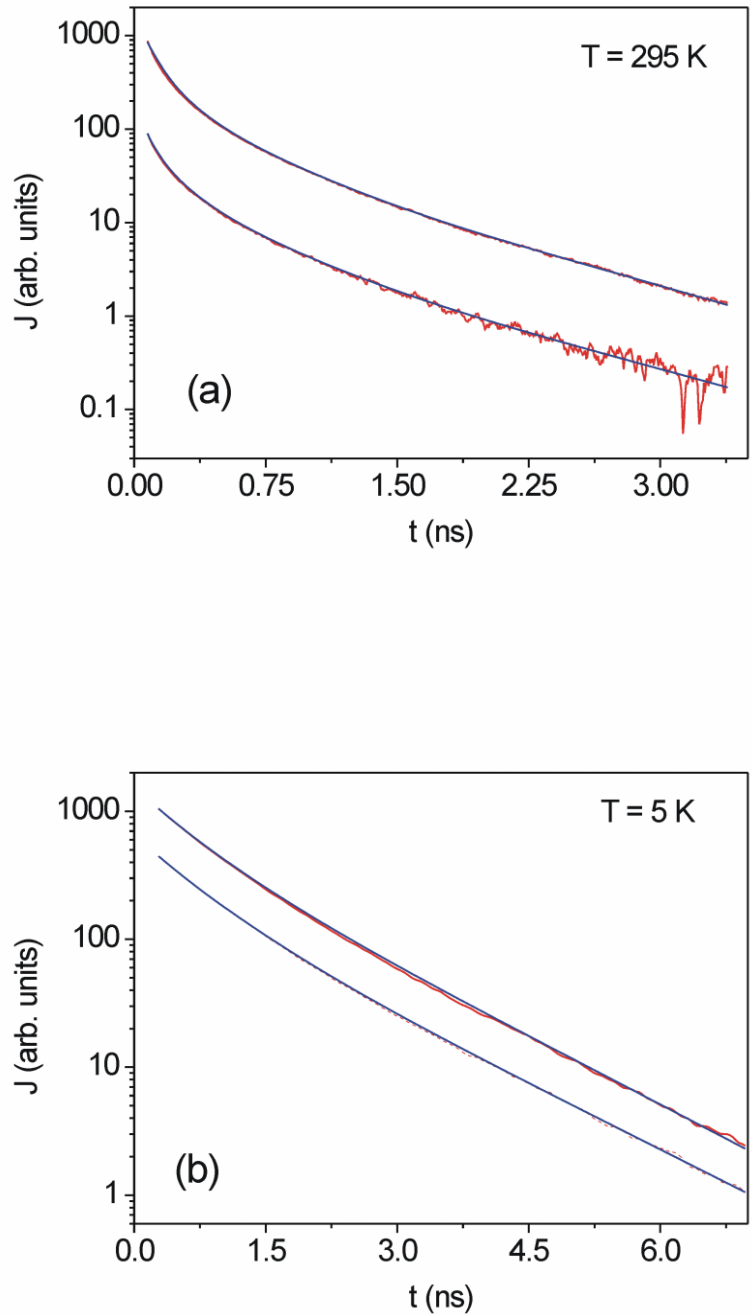

**Fig. 4.** PL time decay of polar nitride MQW structure collected at (a) 295K and (b) 5K in the (3.37 eV, 3.90 eV) energy interval. The red line denotes raw signal with the background subtracted, the blue line – noise averaged line. The dominant emission peaks were located within this interval during entire measurement time (see the spectrum in Fig. 3(a)). The laser power was 1000 and 100 μW, and 500 and 200 μW for room and low temperature, respectively. The excitation wavelength was 300 nm.

*3.4.Results and discussion: polar MQWs*

With the used laser power we estimate a photo-generated carrier density in the range of $10^{17} - 10^{18}$ cm$^{-3}$. Thus, at room temperature, in the samples under study, the density of photo-generated carriers is expected to be lower than the equilibrium density of majority carriers coming from Si donors [47]. Therefore, the PL intensity is proportional to the excess density of minority carriers as in the low excitation regime, in accordance to Eq. 19, and a linear dependence is expected, as confirmed by the fit presented in Fig. 5(a). According to Eq. 22, the high value of the liner coefficient confirms important contribution of Auger process in the optical relaxation at high temperatures.

For low temperature T = 5K, the majority carriers are thermalized [47], and their density is small compared to the photo-generated density of the carriers. In such a case the PL intensity is assumed to be proportional to the density of both carrier types, electrons and holes, thus effectively proportional to the square of the new carrier density. Therefore the low temperature PL intensity is plotted as a function of the square root of PL intensity in Fig. 5(b). Both diagrams are scaled to the initial PL intensity, i.e. the following dependence is used:

$$F(y) = \alpha + \beta y + \gamma y^2 \tag{26}$$

where $y = \frac{J}{J(t_o)}$ and $y = \frac{J^{1/2}}{J^{1/2}(t_o)}$ for Fig. 5(a) and Fig. 5(b), respectively. Accordingly, the TRPL signal from the polar MQWs, collected at low temperature (T = 5K), using laser power excitation L = 500 μW and L = 200 μW, are characterized by the square root dependence of the instantaneous PL decay rate on the PL intensity. Thus this diagram confirms that the sample evolves under high excitation regime. The data were used to obtain the values of α, β and γ parameters, denoting relative weight of SRH, radiative and Auger recombination in overall instantaneous PL relaxation rate $r_{PL}$. The PL relaxation rate decreases from 1.29 ns$^{-1}$ down to 0.83 ns$^{-1}$ (Fig. 5(b)). From the linear fit to the noise-free data the value of coefficient α was determined: α = 0.73 ± 0.01 ns$^{-1}$, and the value of parameter A could be determined: *A* = 0.37 ± 0.01 ns$^{-1}$. The value of the parameter β, describing radiative recombination, is: β = 0.54 ± 0.01 ns$^{-1}$. Finally, the value of the Auger-elated recombination coefficient is: γ = 0.02 ± 0.01 ns$^{-1}$, confirming relatively low contribution of the Auger process at low temperatures.

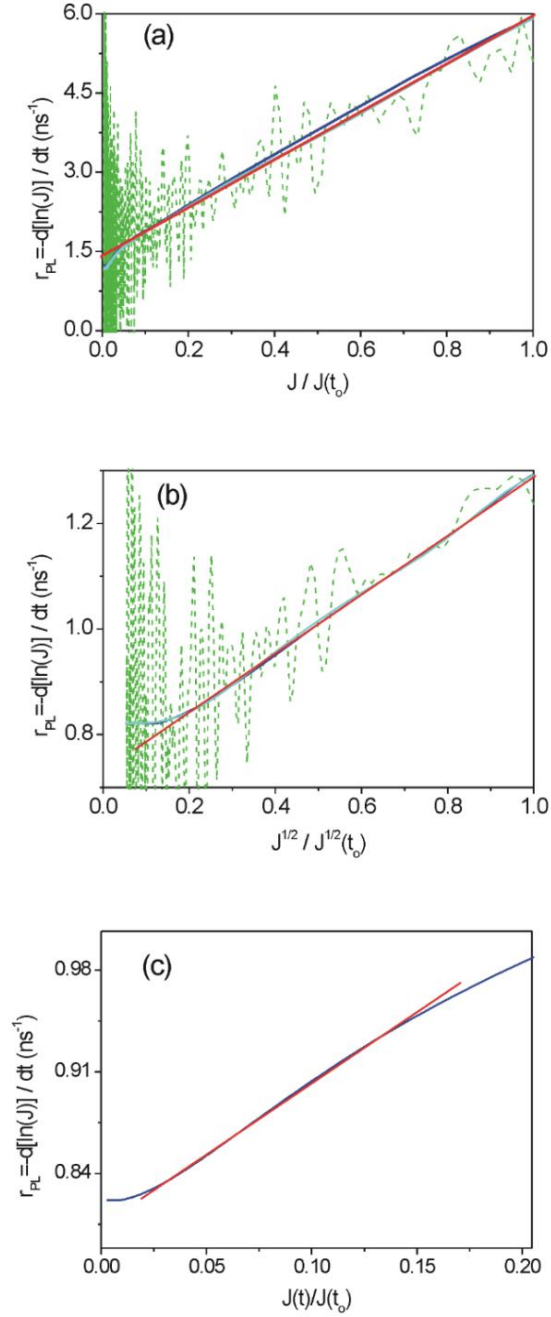

**Fig. 5.** Logarithmic time derivative as a function of the normalized PL intensity of polar nitride MQW structure at: (a) room temperature, (b) and (c) – at 5K. The diagrams (a) and (b) presents noisy signal as dashed green line. The room temperature diagram (a) presents the denoised data for laser power $L$ = 1000 μW (blue line) and $L$ = 100 μW (cyan line) and linear approximation (red line) in function of PL intensity. The low temperature diagram (b) presents the denoised data for laser power $L$ = 500 μW (blue line) and $L$ = 200 μW (cyan line) and linear approximation (red line) in function of square root of PL intensity. The diagram (c) presents the low intensity part of diagram (b), corresponding to Process II of Fig 1(a), in function of PL intensity. The finite difference quotients of raw noisy data for higher laser power (green dashed lines) are also shown to give a visual idea of the accuracy of the smoothing method.

The relative ratio of the nonlinear to linear recombination rate $\left(\frac{\beta+\gamma}{\propto}\right)$ decreases from 0.768 to zero for long times i.e. for small $J$. As it is shown for low temperatures the tri-molecular (Auger) recombination is negligible. Thus it may be concluded that practically entire nonlinear recombination is due to radiative recombination. The radiative recombination part $\left(\frac{\beta}{\propto+\beta+\gamma}\right)$ may be therefore estimated to be at least 42% of the total excitation energy loss, which is relatively high value.

The low intensity part of the low temperature diagram, presented in Fig 5(c), corresponds to Process II regime (Fig 1(a)). The instantaneous PL relaxation rate may be approximated using Eq. 22. From the linear fit to the dependence on PL intensity, presented in Fig 5(c), the following parameter values are obtained: $A_{eff} = 0.80 \mp 0.01 \, ns^{-1}$, and $B_{eff} = 0.98 \mp 0.01 \, ns^{-1}$. From direct inspection of Fig 5(b), the transition point from Process I to Process II may be identified to $J_{eq}^{1/2}/J^{1/2}(t_o) = 0.2$. The majority carriers density ratio may be obtained via Eq. 8 to be the same value, i.e.: $n_{eq}/n_o = 0.2$. This confirms relatively important role of Auger process.

The value of parameter β could be used to determine the value of parameter B from the relation $\beta = 2Bn_o$. Unfortunately, the determination of the value of the initial electron density $n_o$ is burdened by gross error, related to strong polarization and piezo induced electric fields in GaN/AlN MQWs structure. This well-known Quantum Confined Stark Effect (QCSE) shifts away the holes and electrons reducing their overlap and the radiative recombination efficiency. Additionally, the QW system is located close to the surface (Fig. 2(d)), at which the Fermi level is pinned to broken bond Al state, of its energy located approximately 0.5 eV below the conduction band minimum (CBM). Hence the precise determination of the initial carrier density $n_o$ needs further studies. Possible close related methods include determination of the shift of PL energy caused by screening by free electrons or the investigation of the dependence of PL intensity on laser power at different temperatures which leads to determination of transition from high excitation to low excitation regime.

The estimate of the generated density of electrons may be based on observation that at room temperature the radiative recombination follows the linear regime. Thus the generated electron density is not higher than the equilibrium density of electrons at room temperature, i.e. $n_o < n_{eq} \approx 5.97 \times 10^{18} cm^{-3}$. In this estimate it was assumed that the ionization energy of Si donor is 20 meV. From these data the following estimate for $B$ parameter is obtained: $B > 4.51 \times 10^{-11} s^{-1} cm^{-3}$.

As it is shown in Fig 5(a), the room temperature PL intensity evolves in low excitation regime in which the overall instantaneous PL relaxation rate $r_{PL}$ scales linearly with the independent variable for room temperature measurements. It may be observed that the linear dependence is strictly followed by experimental data for $L = 100$ μW while the higher power data show weak square root behavior. From the fit to the denoised data, the coefficient $\alpha$ denoting the monomolecular process is determined to be: $\alpha = 1.42 \pm 0.01$ ns$^{-1}$, the nonlinear coefficient $\beta = 4.63 \pm 0.03$ ns$^{-1}$, and the tri-molecular coefficient is: $\gamma = -0.14 \pm 0.03$ ns$^{-1}$. The latter coefficient has to be interpreted as the measure of deviation from ABC model as it is expected to vanish.

These data should then be interpreted using the effective coefficient as defined in Eqs 21 and 24. Accordingly for room temperature $\alpha = A_{eff} = A + Bn_{eq} + C_1 n_{eq}^2 = 1.42\ ns^{-1}$, which could be compared to the low temperature value $A = 0.37 \pm 0.01$ ns$^{-1}$. This may be therefore set as reference only, leaving the remaining difference equal to $1.06 \pm 0.03$ ns$^{-1}$ as a contribution from radiative recombination and Auger processes. From comparison to the low temperature B parameter values it follows that a considerable contribution may stem from Auger process that is confirmed by high nonlinear value of coefficient $B_{eff}$. The larger contribution of Auger process at room temperature stems from relatively large contribution of equilibrium carrier density (quadratic) according to Eqs. 4 and 1.

The time evolution of the optical emission may be presented in more intuitive way by plots of the instantaneous PL recombination rate $r_{PL}$ vs time. The diagrams presenting such evolutions are shown in Figure 6 (a) and (b). As it could be seen the rate decreases sharply in the initial part of the evolution, confirming the important role of non-linear contributions both for low and room temperature PL decay. No drastic dependence on the laser excitation power was observed, indicating identical evolution of excitation. Naturally, the asymptotic evolution is strictly linear as evidenced by the horizontal line in the final stage of all decays. The nonlinear part of the evolution of PL decay, needed to attain the linear evolution at the final stage of the decay is different: the slope becomes horizontal after 2.5 ns at room temperature and after 5 ns at 5 K. Thus temperature dependent contribution plays an important role. This is confirmed by the different asymptotic rate which is $\lim_{t \to \infty} r_{PL} =$ $0.82\ ns^{-1}$ for T = 5 K and is $\lim_{t\to\infty} r_{PL} = 1.23\ ns^{-1}$ at room temperature. Thus the temperature dependent term contributes to the asymptotic relaxation rate indicating a considerable contribution of non-radiative recombination in these systems at room temperature.

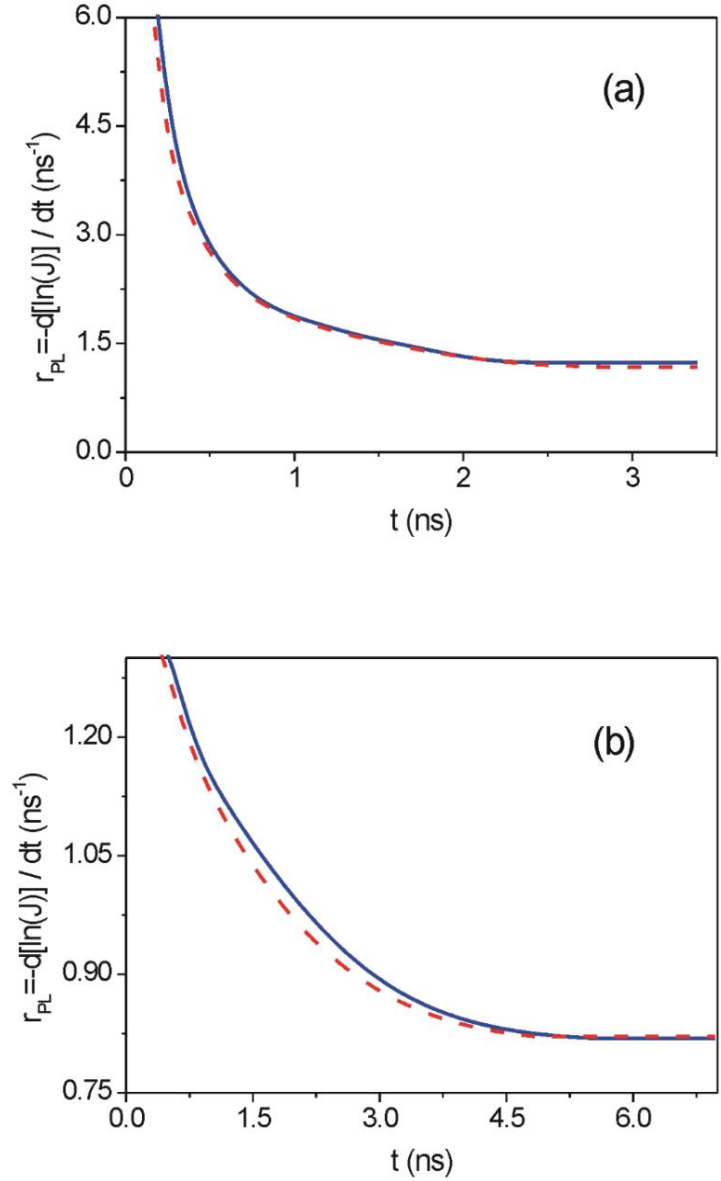

**Fig. 6.** The instantaneous PL recombination rate $r_{PL}$ in function of time in polar nitride MQW structure at: (a) room temperature, (b) at 5K. The room temperature diagram (a) presents the denoised data for laser power $L$ = 1000 μW (solid blue line) and $L$ = 100 μW (red dashed line). The low temperature diagram (b) presents the denoised data for laser power $L$ = 500 μW (blue solid line) and $L$ = 200 μW (red dashed line).

It has to be stressed out that these diagrams allow to obtain valuable insight into the time evolution of the optical relaxation in the MQWs systems. Additionally, the asymptotic relaxation (linear) rate could be determined with high precision.

### 3.5. Results and discussion: non-polar MQWs

The polar MQWs are affected by built-in and piezo-induced internal electric fields affecting electrons and holes. Therefore it is also interesting to investigate non-polar MQWs, which are free of these effects. For comparison we present room and low temperature TRPL data obtained from the structure consisting of 50 periods of 3.0-nm-thick GaN wells and 22.5-nm-thick $Al_{0.26}Ga_{0.74}N$ barriers. The time evolution is presented in Fig. 7.

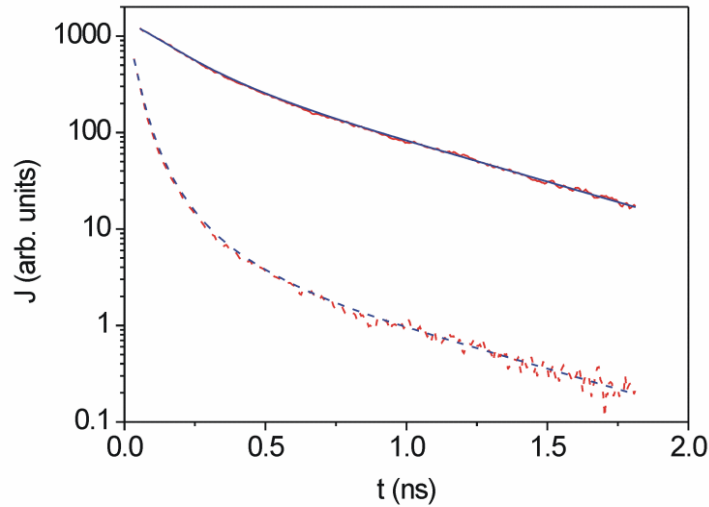

**Fig. 7.** PL time decay of non-polar nitride MQW structure collected at 300K (dashed lines) and 5K (solid lines) in the (3.374 eV, 3.898 eV) energy interval. The red line denotes raw signal with the background subtracted, the blue line – noise averaged line. The dominant emission peaks were located within this interval during entire measurement time. The laser power was 500 µW and the excitation wavelength was 300 nm.

From these data it follows again that the long time evolution is exponential, similar to the data obtained for narrow polar QWs. Thus for 1 nm wells, the electric field does not play important role. As before, the analysis could be made using Eqs 11 and 22 so that the logarithmic time derivative of PL intensity can be obtained after application of appropriate smoothing procedure. The results are presented in Fig. 8.

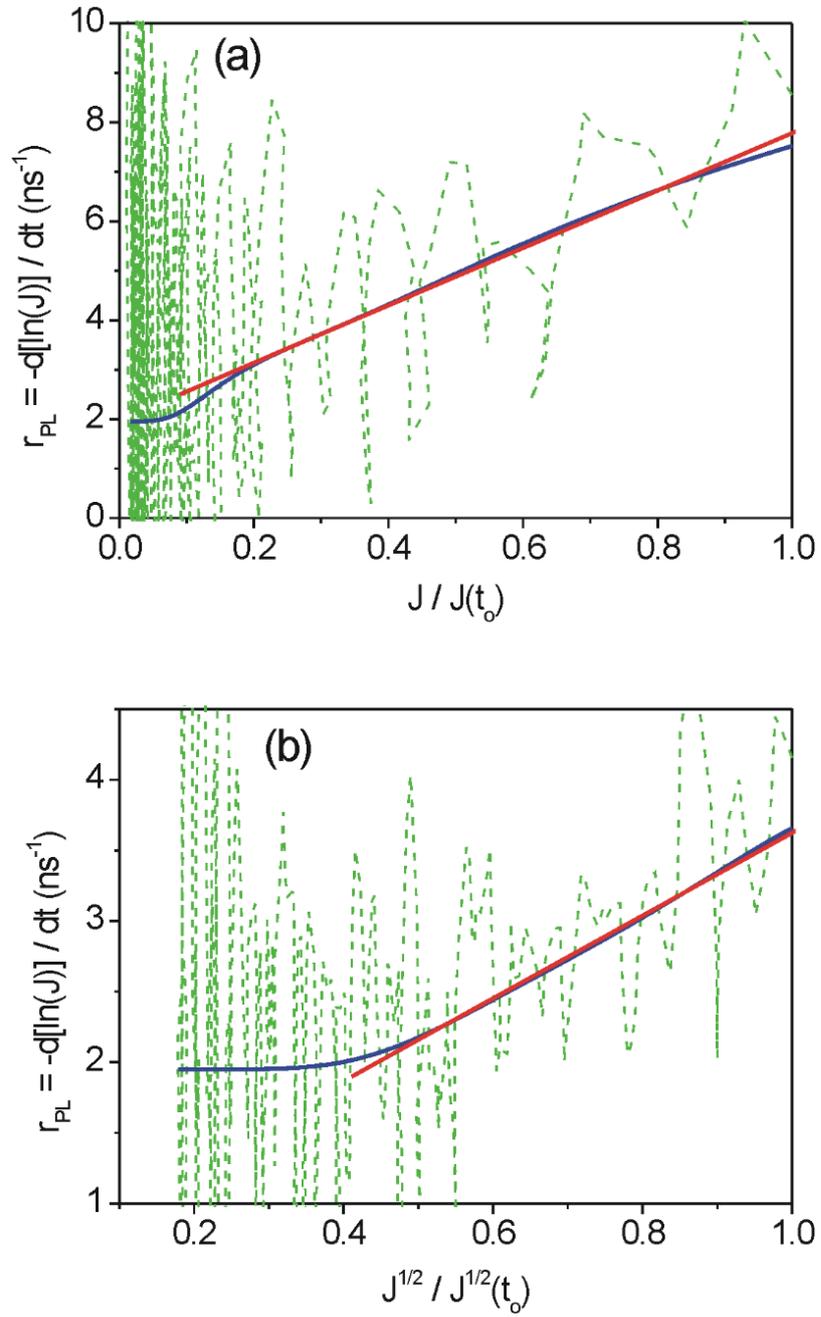

**Fig. 8.** Logarithmic time derivative as a function of the normalized PL intensity of non-polar nitride MQW structure at (a) room temperature and (b) at 6 K: denoised data (blue line) and linear approximation (red line). The finite difference quotients of raw noisy data (green dashed lines) are also shown to give a visual idea of the accuracy of the smoothing method.

The low temperature TRPL evolution from non-polar wells also follows quadratic dependence of PL intensity on excitation power, thus Eqs. 11 and 14 may be applied. The instantaneous PL decay rate is higher than in the case of polar QWs: it decreases from 3.64 ns$^{-1}$ to 1.96 ns$^{-1}$. The value of the parameter α is: α = 1.04 ± 0.01 ns$^{-1}$, i.e. it is slightly higher than the low temperature value for the case of polar QWs. The value of parameter *A* is then equal to *A* = 0.52 ± 0.01 ns$^{-1}$. This difference needs more detailed study, as it was suggested recently that SRH recombination may depend on the electric field [37,38]. The value of parameter β is: β = 1.92 ± 0.03 ns$^{-1}$. Finally, the value of the Auger recombination coefficient is: γ = 0.72 ± 0.02 ns$^{-1}$. Thus these data differ from the previous ones, which may be partially related to the absence of electric field in non-polar wells. The value of parameter B may be assessed as before: $B > 1.05 \times 10^{-10} cm^{-3}$.

Similarly to polar wells, the time evolution of the optical emission is presented in more intuitive way by plots of the instantaneous PL recombination rate $r_{PL}$ vs time in the diagrams shown in Figure 9 (a) and (b). As it is shown, the nonlinear evolution dominates the initial PL decay of the non-polar system for both room temperature and at 6 K. As it is shown the asymptotic rate is $\lim_{t \to \infty} r_{PL} = 1.95\ ns^{-1}$ for 6 K and $\lim_{t \to \infty} r_{PL} = 1.95\ ns^{-1}$ at room temperature, i.e. they are essentially identical. That indicates no temperature dependent contribution in the asymptotic regime. Thus, the recombination is only radiative and temperature dependent SRH recombination is negligible. That is drastically different from the previous polar MQWs system, indicating that the structure of defects is different in these two cases.

It is also worth to note that the initial evolution is different. As it is shown in Fig 9, the room temperature evolution is drastically faster, both the initial stage of the evolution which is terminated in 0.1 ns, and not visible in Fig 9a. The low temperature evolution is much slower, taking about 0.15 ns, as visualized in Fig 9b. The initial rate is much higher $r_{PL} \cong 30\ ns^{-1}$ for room temperature and $r_{PL} \cong 4\ ns^{-1}$ at 6K. Thus temperature dependent processes play important role at high excitations.

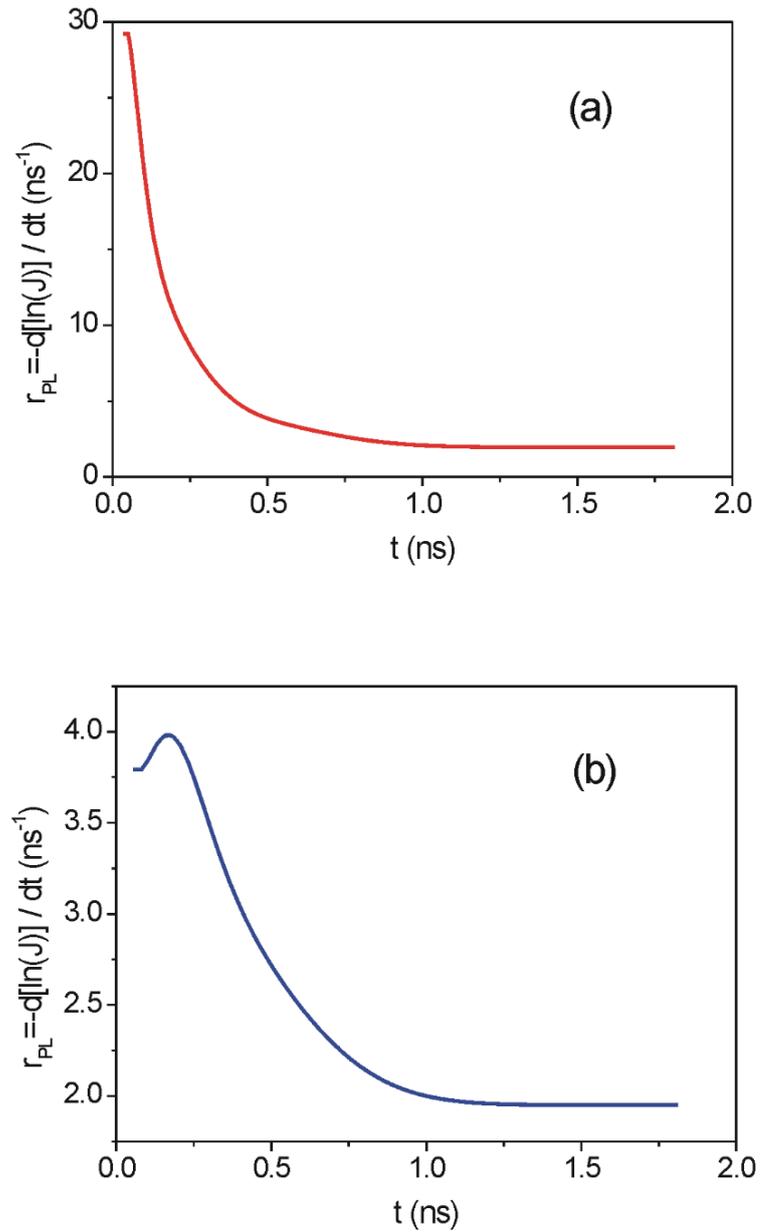

**Fig. 9.** The instantaneous PL recombination rate $r_{PL}$ in function of time in non-polar nitride MQW structure at: (a) room temperature, (b) at 5K. The laser power was 500 μW and the excitation wavelength was 300 nm.

It is instructive to compare diagrams plotted in Fig 8 and 9 and the derived data. As it is shown in Fig 8 and 9 the asymptotic recombination rate is equal to $\lim_{t \to \infty} r_{PL} = 1.95 \, ns^{-1}$. Nevertheless this is product of identical regime in both cases, i.e. linear. As shown in Fig 8 the asymptotic regime differs drastically from the basic evolution. This sort of information cannot be drawn from Fig. 9, where no trace of the difference in recombination modes could be detected. Therefore the new diagrams, such as in Figs 5 and 8 create new valuable tool for optical investigations providing the information, which could not be obtained in traditional analysis of time dependent plots.

From these data, TRPL evolution of narrow polar and non-polar QWs shows remarkable similarity. Note however that despite identical growth conditions, due to the different crystallographic orientation, the point defect and also extended defect densities may be different inducing some difference in the PL data.

It has to be noted also that the obtained values may be burdened by errors stemming not only from the estimate of the initial carrier density but also from other physical effects, like charge carrier diffusion, polarization charge or excitation pulse length [48-51]. These factors will be studied and reported in the following publications.

## 4. Summary and conclusions

We have presented a new method of analysis of TRPL data, based on the instantaneous PL recombination rate $r_{PL}$, i.e. the dependence of logarithmic time derivative of the PL intensity in function of the PL intensity or time. The transformation of the TRPL data is exact, avoiding any approximations. The application of the method is universal, it could be used to analyze the PL decay kinetics of any system, including semiconducting and molecular structures. The method provides graphic representation of the results allowing identification of the mono, bi-, tri-molecular and higher order recombination processes. Therefore the method could be used to determine standard ABC parameters of excess-carrier-density time evolution. Successful application of the method requires two conditions to be met: efficient noise reduction scheme and the precise determination of the zero signal level. The procedures describing how to attain these two goals were also presented.

The presented method was used to identify the value of the monomolecular A parameter, and assess the contribution of higher order processes, including Auger recombination in the exemplary polar and non-polar GaN/AlN MQW systems. Potentially, the method may precisely verify various models of the efficiency droop in the nitride devices.

In general, instantaneous PL recombination rate $r_{PL}$ changes drastically from the initial values proving the dominant contribution of non-linear terms. These terms are temperature dependent, thus indicating that non-radiative recombination prevails at high excitation. The final stage is strictly linear, showing consistent behavior. The differences in asymptotic regime are also observed: a considerable temperature contribution exists for polar MQWs while for non-polar it is absent. That indicates the considerable contribution of SRH recombination for polar MQWs while these effects are absent for non-polar MQWs systems.

The obtained low temperature (T = 5K) values of the monomolecular recombination A parameter are finite, equal to: $A = 0.37 \pm 0.01$ ns$^{-1}$ and $A = 0.52 \pm 0.01$ ns$^{-1}$ for polar and non-polar MQWs, respectively. The respective monomolecular recombination times are: $\tau_{m-pol} = 2.73\ ns$ and $\tau_{m-nonpol} = 1.93\ ns$, indicating that the monomolecular recombination is not negligible also at low temperature. This result shows also that recombination in polar material is slower.

The low temperature data confirm relatively small contribution of Auger recombination at low excitations. Room temperature data indicate the important contribution of Auger processes to non-radiative recombination.

**Acknowledgements:**

This work was supported by the Polish National Science Center under grant number DEC-2015/19/B/ST5/02136, and by the European Union within the European Regional Development Fund through the Innovative Economy grant POIG.01.01.0200-108/09. The TEM study was carried out at the CNBCh UW, established within the project co-financed by EU from the European Regional Development Fund under the Operational Programme Innovative Economy, 2007 – 2013 and NCBR Panda2 501-D312-56-0000002.

**References:**

[1] G.F.J. Garlic, Luminescence, in: G. Flugge (Ed.), Light and Matter II, Springer-Verlag, Berlin-Heidelberg 1958, pp. 1-128.
[2] H.B. Bebb, E.W. Williams, Photoluminescence I. Theory, in: R.K. Willardson, A. C. Beer (Eds.), Semiconductors and Semimetals, Academic Press, New York 1972, Vol. 8, pp. 181-320.

[3] V. P. Gribkovskii, Theory of Luminescence, in: D. R. Vij (Ed.), Luminescence of Solids, Plenum Press, New York 1998, pp. 1-43.

[4] I. Pelant, J. Valenta, Luminescence Spectroscopy of Semiconductors, Oxford Univ. Press, Oxford 2012.

[5] M. Sze, Physics of Semiconductor Devices, Wiley, New York 2001.

[6] W. Shockley, W.T. Read, Jr., Statistics of the recombinations of holes and electrons, Phys. Rev. 87 (1952) 835-842. https://doi.org/10.1103/PhysRev.87.835 .

[7] G. H. Wannier, The structure of electronic excitation levels in insulating crystals, Phys. Rev. 52 (1937) 191-197. https://doi.org/10.1103/PhysRev.52.191 .

[8]. N. F. Mott, Conduction in polar crystals. II. The conduction band and ultra-violet absorption of alkali-halide crystals, Trans. Faraday Soc. 34 (1938) 500-506. https://doi.org/10.1039/TF9383400500 .

[9] M. H. Anderson, J. R. Ensher, M. R. Matthews, C. E. Wieman, and E. A. Cornell, Observation of Bose-Einstein condensation in a dilute atomic vapor, Science, 269 (1995) 198-201. https://doi.org/10.1126/science.269.5221.198 .

[10] S. Schmitt-Rink, D.S. Chemla, D.A.B. Miller, Theory of transient excitonic optical nonlinearities in semiconductor quantum-well structures, Phys. Rev. B. 32 (1985) 6601-6609. https://doi.org/10.1103/PhysRevB.32.6601

[11] E. Zielinski, F. Keppler, S. Hausser, M.H. Pilkuhn, R. Sauer, W.T. Tsang, Optical gain and loss processes in GaInAs/InP MQW laser structures, IEEE J. Quant. Electron. 25 (1989) 1407-1416. https://doi.org/10.1109/3.29276 .

[12] M.F. Schubert, S. Chhajed, J.K. Kim, E.F. Schubert, D.D. Koleske, M.H. Crawford, S.R. Lee, A.J. Fischer, G. Thaler, M.A. Banas, Effect of dislocation density on efficiency droop in GaInN∕GaN light-emitting diodes, Appl. Phys. Lett. 91 (2007) 231114-1-3. https://doi.org/10.1063/1.2822442 .

[13] Q. Dai, Q. Shan, J. Weng, S. Chhajed, J. Cho, E.F. Schubert, M.H. Crawford, D. D. Koleske, M.-H. Kim, Y. Park, Carrier recombination mechanisms and efficiency droop in GaInN/GaN light-emitting diodes, Appl. Phys. Lett. 97 (2010) 133507-1-3. https://doi.org/10.1063/1.3493654 .

[14] M.G. Brik, S. Mahlik, D. Jankowski, P. Strak, K.P. Korona, E. Monroy, S. Krukowski, A. Kaminska, High-pressure studies of radiative recombination processes in semiconductors and rare earth ions doped solids, Jpn. J. Appl. Phys. 56 (2017) 05FA02. https://doi.org/10.7567/JJAP.56.05FA02

[15] F. Wootan, Optical Properties of Solids, Academic Press, New York, 1972.

[16] M. Inokuti, F. Hirayama, Influence of energy transfer by the exchange mechanism on donor luminescence, J. Chem. Phys. 43 (1965) 1978-1989. https://doi.org/10.1063/1.1697063

[17] M. Millot, Z.M. Geballe, K.M. Yu, W. Walukiewicz, R. Jeanloz, Red-green luminescence in indium gallium nitride alloys investigated by high pressure optical spectroscopy, Appl. Phys. Lett. 100 (2012) 162103. https://doi.org/10.1063/1.4704367 .

[18] S. Chichibu, T. Azuhata, T. Sota, S. Nakamura, Spontaneous emission of localized excitons in InGaN single and multiquantum well structures, Appl. Phys. Lett. 69 (1996) 4188-4190. https://doi.org/10.1063/1.116981 .

[19] S. Chichibu, T. Azuhata, T. Sota, S. Nakamura, Luminescences from localized states in InGaN epilayers, Appl. Phys. Lett. 70 (1997) 2822-2824. https://doi.org/10.1063/1.119013 .

[20] S.F. Chichibu, A. Uedono, T. Onuma, B.A. Haskell, A. Chakraborty, T. Koyama, P.T. Fini, S. Keller, S.P. Denbaars, J.S. Speck, U.K. Mishra, S. Nakamura, S. Yamaguchi, S. Kamiyama, H. Amano, I. Akasaki, J. Han, T. Sota, Origin of defect-insensitive emission probability in In-containing (Al,In,Ga)N alloy semiconductors, Nature Mater. 5 (2006) 810-816. https://doi.org/10.1038/nmat1726 .

[21] R.A. Oliver, S.E. Benett, T. Zhu, D.J. Beesley, M.J. Kappers, D.W. Saxey, A. Cerezo, C.J. Humphreys, Microstructural origins of localization in InGaN quantum wells, J. Phys. D: Appl. Phys. 43 (2010) 354003. https://doi.org/10.1088/0022-3727/43/35/354003 .

[22] F. Bernardini, V. Fiorentini, D. Vanderbilt, Polarization-based calculation of the dielectric tensor of polar crystals, Phys. Rev. Lett. 79 (1997) 3958-3961. https://doi.org/10.1103/PhysRevLett.79.3958

[23] A. Kaminska, P. Strak, J. Borysiuk, K. Sobczak, J. Z. Domagala, M. Beeler, E. Grzanka, K. Sakowski, S. Krukowski, E. Monroy, Correlation of optical and structural properties of GaN/AlN multi-quantum wells - *ab initio* and experimental study, J. Appl. Phys. 119 (2016) 015703-1-10. https://doi.org/10.1063/1.4939595 .

[24] P. Strak, P. Kempisty, K. Sakowski, A. Kaminska, D. Jankowski, K. P. Korona, K. Sobczak, J. Borysiuk, M. Beeler, E. Grzanka, E. Monroy, S. Krukowski, *Ab initio* and experimental studies of polarization and polarization related fields in nitrides and nitride structures, AIP Advances 7 (2017) 015027-1-26. http://dx.doi.org/10.1063/1.4974249

[25] P. Strak, P. Kempisty, K. Sakowski, S. Krukowski Doping effects in InN/GaN short-period quantum well structures - theoretical studies based on density functional methods, J. Cryst. Growth 401 (2014) 652-656. http://dx.doi.org/10.1016/j.jcrysgro.2014.01.069 .

[26] A. Suchocki, J.M. Langer, Auger effect in the $Mn^{2+}$ luminescence of CdF2:(Mn,Y) crystals, Phys. Rev. B 39 (1989). 7905-7916 . https://doi.org/10.1103/PhysRevB.39.7905 .

[27] S. Hausser, G. Fuchs, A. Hangleiter, K. Streubel, W.T. Tsang, Auger recombination in bulk and quantum well InGaAs, Appl. Phys. Lett. 56, (1990) 913-915. https://doi.org/10.1063/1.103175 .

[28] J. Cho, E.F. Schubert, J.K. Kim, Efficiency droop in light-emitting diodes: Challenges and countermeasures, Laser Photon. Rev. 7 (2013) 408-421. https://doi.org/10.1002/lpor.201200025 .

[29] M. Shahmohammadi, W. Liu, G. Rossbach, L. Lahourcade, A. Dussaigne, C. Bougerol, R. Butte, N. Grandjean, B. Deveaud, G. Jacopin, Enhancement of Auger recombination induced by carrier localization in InGaN/GaN quantum wells, Phys. Rev. B 95 (2017) 125314-1-10. https://doi.org/10.1103/PhysRevB.95.125314 .

[30] M. R. Krames, G. Christenson, D. Collins, L. W. Cook, M. G. Craford, A. Edwards, R. M. Fletcher, N. F. Gardner, W. K. Goetz, W. R. Imler, E. Johnson, R. S. Kern, R. Khare, F. A. Kish, C. Lowery, M. J. Ludowise, R. Mann, M. Maranowski, S. A. Maranowski, P. S. Martin, J. O'Shea, S. L. Rudaz, D.A. Steigerwald, J. Thompson, J.J. Wierer, J. Yu, D. Basile, Y.-L. Chang, G. Hasnain, M. Heuschen, K.P. Killeen, C.P. Kocot, S. Lester, J.N. Miller, G.O. Mueller, R. Mueller-Mach, S.J. Rosner, R.P. Schneider, T. Takeuchi, T.S. Tan, High-brightness AlGaInN light-emitting diodes Proc. SPIE 3938, (2000) 2-12. https://doi.org/10.1117/12.382822

[31] T. Mukai, M. Yamada, S. Nakamura, Characteristics of InGaN-Based UV/Blue/Green/Amber/Red Light-Emitting Diodes, Jpn. J. Appl. Phys. 38 (1999) 3976-3981. https://doi.org/10.1143/JJAP.38.3976 .

[32] J. Piprek, F. Roemer, B. Witzigmann, On the uncertainty of the Auger recombination coefficient extracted from InGaN/GaN light-emitting diode efficiency droop measurements, Appl. Phys. Lett. **106**, (2015) 101101-1-4. http://dx.doi.org/10.1063/1.4914833 .

[33] T.J. Badcock, M. Ali, T. Zhu, M. Pristovsek, R.A. Oliver, A.J. Shields, Radiative recombination mechanisms in polar and non-polar InGaN/GaN quantum well LED structures, Appl. Phys. Lett. 109 (2016) 151110-1-5. https://doi.org/10.1063/1.4964842 .

[34] G. Verzelesi, D. Saguatti, M. Meneghini, F. Bartazzi, M. Goano, G. Meneghesso, E. Zanoni, Efficiency droop in InGaN/GaN blue light-emitting diodes: Physical mechanisms and remedies, J. Appl. Phys. 114 (2013) 071101-1-14. https://doi.org/10.1063/1.4816434 .

[35] H. Amano, Development of the Nitride Based UV/DUV LEDs, in: B. Gil (Ed.), III-Nitride Semiconductors and Their Modern Devices, Oxford Univ. Press, Oxford, 2013, pp. 1-17.

[36] Y.C. Shen, G.O. Mueller, S. Watanabe, N.F. Gardner, A. Munkholm, M.R. Krames, Auger recombination in InGaN measured by photoluminescence, Appl. Phys. Let. 91 (2007) 141101-1-3. https://doi.org/10.1063/1.2785135 .

[37] A. David, N.G. Young, C. A. Hurni, M.D. Craven, All-optical measurements of carrier dynamics in bulk-GaN LEDs: Beyond the ABC approximation, Appl. Phys. Lett. 110 (2017) 253504. https://doi.org/10.1063/1.4986908 .

[38] A. David, C. A. Hurni, N.G. Young, M.D. Craven, Field-assisted Shockley-Read-Hall recombinations in III-nitride quantum wells, Appl. Phys. Lett. 111 (2017) 233501. https://doi.org/10.1063/1.5003112 .

[39] M.P. Halsall, J.E. Nicholls, J.J. Davies, B. Cockayne, P.J. Wright, CdS/CdSe intrinsic Stark superlattices, J. Appl. Phys. 71 (1992) 907. https://doi.org/10.1063/1.351312 .

[40] T. Takeuchi, C. Wetzel, S. Yamaguchi, H. Sakai, H. Amano, I. Akasaki, Y. Kaneko, S. Nakagawa, Y. Yamaoka, N. Yamada, Determination of piezoelectric fields in strained GaInN quantum wells using the quantum-confined Stark effect, Appl. Phys. Lett. **73**, 1691 (1998). https://doi.org/10.1063/1.122247 .

[41] A. Kaminska, D. Jankowski, P. Strak, K.P. Korona, M. Beeler, K. Sakowski, E. Grzanka, J. Borysiuk, K. Sobczak, E. Monroy, S. Krukowski, High pressure and time resolved studies of optical properties of n-type doped GaN/AlN multi-quantum wells – experimental and theoretical analysis, J. Appl. Phys. 120 (2016) 095705-1-9. https://doi.org/10.1063/1.4962282

[42] E. Faulques, J. Wery, S. Lefrant, V.G. Ivanov, G. Jonusauskas, Transient photoluminescence of para-hexaphenyl layers, Phys. Rev B 65 (2002) 212202-1-4. https://doi.org/10.1103/PhysRevB.65.212202

[43] J.S. Speck, S.F. Chichibu, Nonpolar and semipolar group III nitride-based materials, MRS Bulletin, 34 (2009) 304-312. https://doi.org/10.1557/mrs2009.91 .

[44] C.B. Lim, A. Ajay, E. Monroy, Gallium kinetics on *m*-plane GaN, Appl. Phys. Lett. 111 (2017) 022101-1-5. https://doi.org/10.1063/1.4993570 .

[45] C. B. Lim, M. Beeler, A. Ajay, J. Lähnemann, E. Bellet-Amalric, C. Bougerol, E. Monroy, Intersubband transitions in nonpolar GaN/Al(Ga)N heterostructures in the short- and mid-wavelength infrared regions, J. Appl. Phys. 118 (2015) 014309-1-8. https://doi.org/10.1063/1.4926423 .

[46] Ghkonrad, Python command-line application for noise reduction of time-resolved photoluminescence experimental measurements. https://github.com/ghkonrad/logpli .

[47] V. Bougrov, M.E. Levinshtein, S.L. Rumyantsev, A. Zubrilov, in: M.E. Levinshtein, S.L. Rumyantsev, M.S. Shur (Eds.), Properties of Advanced Semiconductor Materials GaN, AlN, InN, BN, SiC, SiGe, Wiley, New York, 2001, p. 1-30.

[48] M. Maiberg, R. Scheer, Theoretical study of time-resolved luminescence in semiconductors. I. Decay from the steady state, J. Appl. Phys. 116 (2014) 123710-1-12. https://doi.org/10.1063/1.4896483 .

[49] M. Maiberg, R. Scheer, Theoretical study of time-resolved luminescence in semiconductors. II. Pulsed excitation, J. Appl. Phys. 116 (2014) 123711-1-11. https://doi.org/10.1063/1.4896484 .

[50] M. Maiberg, T. Holscher, S. Zahedi-Azad, and R. Scheer, Theoretical study of time-resolved luminescence in semiconductors. III. Trap states in the band gap, J. Appl. Phys. 118 (2015) 1205701-1-10. https://doi.org/10.1063/1.4929877 .

[51] M. Maiberg, F. Beltram, M. Muller, R. Scheer, Theoretical study of time-resolved luminescence in semiconductors. IV. Lateral inhomogeneities, J. Appl. Phys. 121 (2017) 085703-1-14. https://doi.org/10.1063/1.4976102 .